\documentclass[pra,aps,twocolumn,amsmath,amssymb,superscriptaddress]{revtex4-1}
\usepackage{epsfig,amsmath}
\usepackage{subfigure}
\usepackage{graphicx}
\usepackage{dcolumn}
\usepackage{stmaryrd}
\usepackage{mathrsfs}
\usepackage{pifont}
\usepackage{amsthm}
\usepackage{amssymb}
\usepackage{bm}
\usepackage{latexsym}
\usepackage[colorlinks=true,linkcolor=blue,citecolor=blue]{hyperref}
\usepackage{color}
\usepackage{epstopdf}
\usepackage[ruled]{algorithm2e}
\usepackage{lipsum}

\begin{document}

\title{Efficient nonclassical state preparation via generalized parity measurement}

\author{Chen-yi Zhang}
\affiliation{School of Physics, Zhejiang University, Hangzhou 310027, Zhejiang, China}

\author{Jun Jing}
\email{Contact author: jingjun@zju.edu.cn}
\affiliation{School of Physics, Zhejiang University, Hangzhou 310027, Zhejiang, China}

\date{\today}

\begin{abstract}
Nonclassical states of bosonic modes, especially the large number states, are valuable resources for quantum information processing and quantum metrology. It is however intricate to generate a desired Fock state of bosonic systems by unitary protocols due to their uniform energy spectrum. We here propose a nonunitary protocol that is based on the resonant Jaynes-Cummings interaction of the bosonic mode with an ancillary two-level atom and sequential projective measurements on the atom. Using the generalized parity-measurement operator constructed by several rounds of free evolution with stepwise halved intervals and measurement, we can efficiently filter out the unwanted population and push the target resonator conditionally toward the desired Fock state. In the ideal situation, a Fock state $|n_t\approx2000\rangle$ can be prepared with a fidelity over $98\%$ using only eight rounds of measurements. Under qubit dissipation and dephasing and cavity decay in the current circuit-QED platforms, a Fock state $|n_t\approx100\rangle$ can be prepared with a fidelity of about $80\%$ by six measurements. It is found that the number of measurement rounds for preparing a large Fock state $|n_t\rangle$ scales roughly as $\log_2\sqrt{n_t}$, which is similar to the number of ancillary qubits required in the state preparation via the quantum phase estimation algorithm and yet costs much less in gate operations. Our protocol can also be used to prepare a large Dicke state $|J\simeq1000,0\rangle$ of a spin ensemble with a sufficiently high fidelity by less than six measurements. It is qualified by the quantum Fisher information approaching the Heisenberg scaling in sensing the rotation phase along the $x$ axis.
\end{abstract}

\maketitle

\section{Introduction}

Quantum measurement has a long history serving as a fundamental tool for controlling and manipulating quantum states. First rigorously analyzed by Misra and Sudarshan in 1977~\cite{JMP1977Misra}, the sufficiently frequent projective measurements can ``freeze'' the system in the initial state, which is known as the quantum Zeno effect (QZE)~\cite{AnnaP1997D,PRA1990Itano,PRA2018Zhang}. Closely associated with QZE, a purification protocol was proposed by performing a series of frequent measurements on a system to manipulate the dynamics of another system that is coupled to it~\cite{PRL2003Nakazato,PRANakazato2004}. Following these pioneer works, numerous efforts have been devoted to exploiting Zeno-like measurements for various quantum control tasks, such as entanglement generation~\cite{PRA2004Wu,NJP2009Vacanti,PRA2012Qiu,PRA2011Busch}, ground-state preparation~\cite{PRB2011Li,PRB2020Puebla,PRA2016Pyshkin,PRA2021Yan,PRA2022Konar}, nuclear spin polarization~\cite{JPA2011Wu,PRA2022Jin}, and quantum battery charging~\cite{PRApplied2023Yan}.

More than merely exploiting the frequent measurements on the ancillary system, one can accelerate the state preparation by incorporating appropriate gate operations into the joint free evolution. A widely adopted protocol leverages the quantum phase estimation (QPE) algorithm to prepare the eigenstates $|u\rangle$ of a unitary operator $U|u\rangle=e^{i2\pi\varphi}|u\rangle$ with eigenvalue $e^{i2\pi\varphi}$~\cite{PRL1999Abrams,AnnaP2003Kitaev}. In QPE, the target system of several qubits is coupled to a register of $N$ ancillary qubits. By performing $2^N$ controlled unitary operations, the eigenphase $\varphi$ of the unitary operator is encoded into the ancillary register~\cite{NielsenChuang2010}. Then after an inverse quantum Fourier transform and the subsequent measurements on the ancillary qubits, the target system is collapsed by the algorithm to an eigenstate of the unitary operator with a probability determined by its initial overlap with the resulting state. The eigen-phase can be estimated as a binary decimal:
\begin{equation}\label{phase}
 \varphi=0.\varphi_1\varphi_2\ldots=\sum_{l=1}^N\varphi_l2^{-l},
\end{equation}
where $\varphi_l$ is the measurement outcome of the $l$th ancillary qubit, either zero or one.

Alternatively, one can implement the phase estimation using a single ancillary qubit by iterating a Ramsey-type circuit~\cite{PhysRev1950Ramsey,PRL2014Asadian,PRX2018Fluhmann}. At the beginning of the $k$th round of protocols~\cite{arXiv1995kitaev,NPJ2019Brien,PRA2025Jin} with $1\leq k\leq N$, the ancillary qubit is reset as $(|0\rangle+|1\rangle)/\sqrt{2}$ via a Hadamard gate. Then $2^{k-1}$ controlled-$U$ operations are applied to the target system, followed by a conditional single-qubit rotation along the $z$ axis $R_z(a_k)=|0\rangle\langle 0|+e^{i2\pi a_k}|1\rangle\langle 1|$ of the ancillary qubit, where $a_k$ is determined by the outcomes of all preceding measurements. After another Hadamard gate, the state of the ancillary qubit can be read out by a projective measurement. The system under such $N$ rounds of operation can be pushed to the vicinity of a desired or random eigenstate (or a superposition of them) of the estimation operator $U$ with an uncertainty of the estimated eigen-phase $\delta\varphi\propto2^{-N}$. This uncertainty scaling is consistent with Eq.~(\ref{phase}) in a standard QPE~\cite{PRA2016Terhal}. This protocol was used to prepare the Gottesman-Kitaev-Preskill state as an eigenstate of the displacement operator of the bosonic modes~\cite{PRA2017Duivenvoorden,PRA2016Terhal}, the Dicke state of a spin ensemble as an eigenstate of $U=\exp(-iJ_z)$~\cite{PRA2021Wang}, and the large Fock state of the microwave cavity mode that is dispersively coupled to an ancillary transmon qubit~\cite{NP2024Deng}.

In this work, we propose an efficient protocol for generating nonclassical quantum states in a Hilbert space of large dimensionality via effectively constructing a generalized parity measurement (GPM) by optimizing the time intervals between Zeno-like measurements. The required number of measurement rounds exhibits a logarithmic scaling similar to the number of ancillary qubits in the quantum phase estimation algorithm: $N\sim\log_2\sqrt{n_t}$, where $n_t$ indicates the target number state or the excitation number uncertainty $\sqrt{n_t}=|\alpha|$ if the target resonator (bosonic mode) is initially prepared as a coherent state $|\alpha\rangle$. Our protocol exploits the resonant exchange coupling between the target resonator and the ancillary two-level system (TLS) and the projective measurements on the TLS, which is much less complex than the protocols based on QPE in gate operations and holds a much faster preparation speed than the existing GPM protocol based on dispersive coupling~\cite{NP2024Deng,PRA2023Ma}.

The rest of this paper is organized as follows. In Sec.~\ref{GPM}, we briefly review the existing generalized parity measurement based on the dispersive interaction between bosonic mode and ancillary TLS and the projective measurement on the TLS. In Sec.~\ref{EffectiveGPM}, we propose an alternative and more efficient protocol to realize GPM through the resonant exchange interaction between bosonic mode and ancillary TLS and TLS measurement. In Sec.~\ref{Connection}, our GPM is placed in a much broader context with ground-state cooling by measurement. In Sec.~\ref{LargeFock}, we present the results about the generation of a large Fock state in both ideal and realistic situations, which exhibit an approximated logarithmic scaling of the measurement numbers as functions of the target number state. In Secs.~\ref{GPMDicke} and~\ref{Dickeresult}, we extend our protocol to generate the Dicke state of a spin ensemble with a similar GPM operator and demonstrate the protocol performance in both state preparation and quantum metrology, respectively. The conclusion is provided in Sec.~\ref{conclusion}.

\section{Theoretical model}\label{Theory}

\subsection{Generalized parity measurement based on dispersive coupling}\label{GPM}

\begin{figure}[htbp]
\centering
\includegraphics[width=0.90\linewidth]{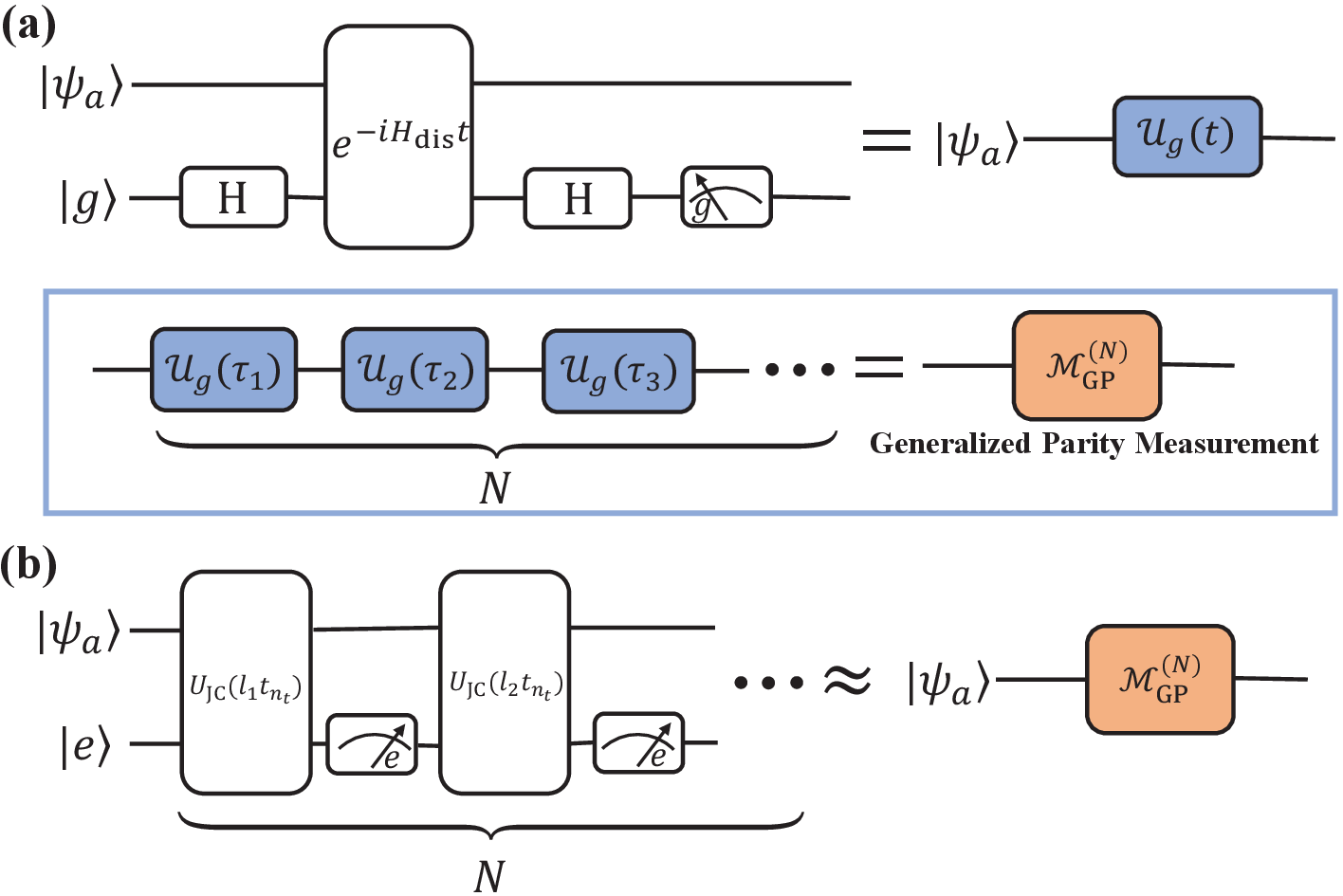}
\caption{(a) Circuit diagram for realizing GPM~\cite{NP2024Deng} through free joint evolution of the dispersively coupled resonator and the ancillary TLS, the Hadamard gates on the TLS, and the projective measurements on the ground state $|g\rangle$ of TLS. (b) Circuit diagram of our protocol for constructing approximated GPM on the target resonator by repeatedly measuring the excited state $|e\rangle$ of the ancillary TLS, which is coupled to the resonator through exchange interaction.}\label{Circuit}
\end{figure}

To efficiently generate a large Fock state $|n_t\rangle$ with $n_t\gg 1$, a generalized parity measurement on the target resonator has been realized by the quantum circuit model in Fig.~\ref{Circuit}(a), which is a standard Ramsey-type sequence~\cite{PhysRev1950Ramsey,NP2024Deng}. In a modern cavity quantum electrodynamics (cavity-QED) platform~\cite{NP2024Deng}, the ancillary TLS (superconducting qubit) is dispersively coupled to the target resonator [a single mode in the three-dimensional (3D) microwave cavity] with a Hamiltonian ($\hbar=1$)
\begin{equation}
H_{\rm dis}=\Delta|e\rangle\langle e|-\chi a^{\dagger}a|e\rangle\langle e|,
\end{equation}
where $\Delta=\omega_q-\omega_a$ is the detuning between qubit and resonator, $\chi$ is the dispersive coupling strength, $|e\rangle$ is the excited state of the qubit, and $a$ ($a^\dagger$) is the annihilation (creation) operator of the bosonic mode. Suppose that the initial state of the resonator is $|\psi_a(0)\rangle=\sum_nC_n|n\rangle$. The qubit is initialized as $(|g\rangle+|e\rangle)/\sqrt{2}$ by a Hadamard gate. After a period of joint evolution governed by $e^{-iH_{\rm dis}\tau}$ with a duration $\tau$, the qubit obtains a phase relevant to the resonator states, and the state of the composite system becomes
\begin{equation}
 \frac{1}{\sqrt{2}}\sum_{n=0}^{\infty}\left[|g\rangle+e^{-i(\Delta-\chi n)\tau}|e\rangle\right]C_n|n\rangle.
\end{equation}
Then another Hadamard gate on the qubit gives rise to the state
\begin{equation}
\begin{aligned}
 &\sum_{n=0}^{\infty}\frac{1}{2}\bigg[\left(1+e^{-i(\Delta-\chi n)\tau}\right)|g\rangle\\
 +&\left(1-e^{-i(\Delta-\chi n)\tau}\right)|e\rangle\bigg]C_n|n\rangle\\
\equiv& \mathcal{U}_g(\tau)|\psi_a(0)\rangle|g\rangle+\mathcal{U}_e(\tau)|\psi_a(0)\rangle|e\rangle,
\end{aligned}
\end{equation}
where
\begin{equation}
\begin{aligned}\label{Ug}
\mathcal{U}_g(\tau)
&=\sum_{n=0}^\infty e^{-i\frac{\Delta-\chi n}{2}\tau}\cos\left(\frac{\Delta-\chi n}{2}\tau\right)|n\rangle\langle n|,\\
\mathcal{U}_e(\tau)
&=i\sum_{n=0}^\infty e^{-i\frac{\Delta-\chi n}{2}\tau}\sin\left(\frac{\Delta-\chi n}{2}\tau\right)|n\rangle\langle n|.
\end{aligned}
\end{equation}
The density matrix of the resonator can be written in a standard Kraus representation
\begin{equation}\label{tilderho}
\tilde{\rho}_a(\tau)=\mathcal{U}_g(\tau)\rho_a(0)\mathcal{U}^{\dagger}_g(\tau)
+\mathcal{U}_e(\tau)\rho_a(0)\mathcal{U}^{\dagger}_e(\tau),
\end{equation}
by tracing out the degree of freedom of the ancillary qubit.

Upon the projective measurement $M_g\equiv|g\rangle\langle g|$ on the ground state of the qubit, the second term in Eq.~(\ref{tilderho}) vanishes. Thus, the quantum circuit [see the upper part of Fig.~\ref{Circuit}(a)] gives rise to a nonunitary evolution operator $\mathcal{U}_g(\tau)$ for the resonator when the measurement outcome turns out to be $|g\rangle$.

Then after $N$ rounds of evolution, Hadamard gates, and measurement, the resonator state becomes
\begin{equation}\label{rhoaN}
\rho_a(N)=\frac{\prod_{k=N}^{1}\mathcal{U}_g(\tau_k)\rho_a(0)\prod_{k=1}^{N}\mathcal{U}^{\dagger}_g(\tau_k)}{P(N)},
\end{equation}
where $\rho_a(0)$ is an arbitrary initial density matrix, $\tau_k$ is the duration of the $k$th round, and $P(N)$ is the success probability of finding the qubit in its ground state for $N$ rounds. With $|n_t\rangle$ the target number state, one can set the detuning to be $\Delta=\chi n_t$ and the duration of the $k$th round as $\tau_k=\pi/(\chi 2^{k-1})$~\cite{NP2024Deng,PRA2023Ma}. In the first round, the nonunitary evolution operator reads
\begin{equation}
\begin{aligned}
\mathcal{U}_g(\tau_1)&=\sum_{n=0}^\infty e^{-i\frac{n_t-n}{2}\pi}\cos\left(\frac{n_t-n}{2}\pi\right)|n\rangle\langle n|\\
&=\sum_{j\geq -\lfloor n_t/2\rfloor}|n_t+2j\rangle\langle n_t+2j|.
\end{aligned}
\end{equation}
It actually defines a parity measurement operator $\mathcal{M}_{\rm P}$ which projects the resonator onto Fock states $|n\rangle$ with an even derivation $|n-n_t|$, i.e., $|n-n_t|/2\in \mathbb{N}$. In the subsequent round, we have
\begin{equation}
\begin{aligned}
&\mathcal{U}_g(\tau_2)=\sum_{n=0}^\infty e^{-i\frac{n_t-n}{4}\pi}\cos\left(\frac{n_t-n}{4}\pi\right)|n\rangle\langle n|\\
=&\sum_{j\geq -\lfloor n_t/4\rfloor}|n_t+4j\rangle\langle n_t+4j|\\
&+\sum_{j\geq -\lfloor (n_t+1)/2\rfloor}e^{-i\frac{2j+1}{4}\pi}\cos\left(\frac{2j+1}{4}\pi\right)\\
&\times|n_t+2j+1\rangle\langle n_t+2j+1|.
\end{aligned}
\end{equation}
Thus, after the first two rounds, the joint nonunitary evolution operator is given by
\begin{equation}
\mathcal{U}_g(\tau_2)\mathcal{U}_g(\tau_1)=\sum_{j\geq -\lfloor n_t/4\rfloor}|n_t+4j\rangle\langle n_t+4j|,
\end{equation}
which can be defined as a generalized parity measurement operator of $2^2$. It projects the resonator onto a subspace of a general parity of $4$, i.e., the subspace spanned by the Fock states $|n\rangle$ satisfying $|n-n_t|/4\in\mathbb{N}$. Then after $k$ rounds, the population of the resonator outside the subspace of a general parity of $2^k$, i.e., the subspace spanned by the Fock states $|n\rangle$ satisfying $|n-n_t|/2^k\in\mathbb{N}$, can be filtered out. On the whole, Eq.~(\ref{rhoaN}) is equivalent to performing a GPM on the resonator:
\begin{equation}\label{raN}
\rho_a(N)=\frac{\mathcal{M}_{\rm GP}^{(N)}\rho_a(0)\mathcal{M}_{\rm GP}^{(N)}}{P(N)},
\end{equation}
where the GPM operator is formally defined as
\begin{equation}\label{MP}
\mathcal{M}_{\rm GP}^{(N)}\equiv\sum_{j\geq -\lfloor n_t/2^N\rfloor}|n_t+2^N j\rangle\langle n_t+2^N j|.
\end{equation}
Clearly $\mathcal{M}_{\rm P}=\mathcal{M}_{\rm GP}^{(1)}$.

We prepare the resonator at the coherent $\rho_a(0)=|\psi_a\rangle\langle\psi_a|$ with $|\psi_a\rangle=|\alpha\rangle=e^{-|\alpha|^2/2}\sum_{n=0}^\infty\alpha^n/\sqrt{n!}|n\rangle$, since it is straightforward to be prepared in experiments (for example, by applying a resonant microwave drive to the resonator). It is assumed that $|\alpha|^2=n_t$ to have the maximum overlap with the target state $|n_t\rangle$, then one can directly find that the number of protocol rounds scales roughly as $N_{\rm Fock}(n_t)\sim\log_2\sqrt{n_t}$ for generating the Fock state $|n_t\gg1\rangle$, since the excitation number uncertainty is $\langle\Delta n\rangle=\sqrt{\langle n^2\rangle-\langle n\rangle^2}=\sqrt{n_t}$. The other choice of $\alpha$ might merely lower the final success probability. This state preparation protocol is similar to those based on QPE to project the system onto the eigenstate with zero eigenphase of the unitary operator $U=e^{-i\pi(n_t-a^{\dagger}a)\chi}$.

\subsection{Efficient GPM in the Jaynes-Cummings model under the resonant condition}\label{EffectiveGPM}

Inspired by the ideas of ground-state cooling~\cite{PRB2011Li,PRA2021Yan} and charging~\cite{PRApplied2023Yan} by measurement, we find that an effective GPM operator similar to Eq.~(\ref{MP}) can be simply constructed by sequentially performing projective measurements $M_e\equiv|e\rangle\langle e|$ on the ancillary TLS that is coupled to the target resonator via the Jaynes-Cummings interaction as shown in Fig.~\ref{Circuit}(b). In the rotating frame with respect to $\omega_aa^{\dagger}a$, the Hamiltonian reads
\begin{equation}
    H_{\rm JC}=\Delta|e\rangle\langle e|+g\left(|e\rangle\langle g|a+|g\rangle\langle e|a^{\dagger}\right),
\end{equation}
where $g$ is the coupling strength. The ancillary two-level system is supposed to be prepared in the excited state $|e\rangle$. Then the initial state of the composite system is given by
\begin{equation}
\rho(0)=|e\rangle\langle e|\otimes\rho_a(0),
\end{equation}
where $\rho_a(0)$ is the initial state of the resonator. After a period of joint evolution by $U_{\rm JC}(\tau)=\exp(-iH_{\rm JC}\tau)$ and tracing out the qubit component, the resonator state can be written as
\begin{equation}
\tilde{\rho}_a(\tau)=\Pi_g(\tau)\rho_a(0)\Pi_g^{\dagger}(\tau)+\Pi_e(\tau)\rho_a(0)\Pi_e^{\dagger}(\tau)
\end{equation}
in the Kraus operator representation, where the Kraus operators $\Pi_{i=e,g}(\tau)$ are
\begin{align}\label{Pie}
\Pi_e(\tau)&=\langle e|U_{\rm JC}(\tau)|e\rangle=\sum_{n=0}^\infty \beta_n(\tau)|n\rangle\langle n|,\\
\Pi_g(\tau)&=\langle g|U_{\rm JC}(\tau)|e\rangle=\sum_{n=0}^\infty \eta_n(\tau)|n+1\rangle\langle n|
\end{align}
with coefficients
\begin{align}\label{betan2}
\beta_n(\tau)&=e^{-i\Delta\tau/2}\left[\cos (\Omega_n\tau)+i\frac{\Delta}{2\Omega_n}\sin(\Omega_n\tau)\right],\\
\eta_n(\tau)&=-ie^{-i\Delta\tau/2}\frac{g\sqrt{n+1}}{\Omega_n}\sin(\Omega_n\tau).
\end{align}
Here $\Omega_n\equiv\sqrt{\Delta^2/4+(n+1)g^2}$ is the $(n+1)$-photon Rabi frequency. If the measurement outcome is $|e\rangle$, the resonator state turns out to be
\begin{equation}
\rho_a(1)=\frac{\Pi_e(\tau)\rho_a(0)\Pi^{\dagger}_e(\tau)}{P_e(1)},
\end{equation}
where $P_e(1)={\rm Tr}[\Pi_e(\tau)\rho_a(0)\Pi^{\dagger}_e(\tau)]$ is the success probability of finding the TLS at $|e\rangle$. $\Pi_e(\tau)$ becomes the nonunitary evolution operator for the circuit in Fig.~\ref{Circuit}(b). To use $\Pi_e(\tau)$ to simulate $\mathcal{U}_g(\tau)$ in Eq.~(\ref{Ug}) and to hold the same target number state $|n_t\rangle$, we let $|\beta_{n_t}(\tau)|=1$. This is equivalent to setting the joint evolution duration as
\begin{equation}\label{taunt}
\tau=lt_{n_t}, \quad t_{n_t}\equiv\frac{\pi}{\Omega_{n_t}}
\end{equation}
with $l\in \mathbb{N}$. By substituting Eq.~(\ref{taunt}) into Eq.~(\ref{betan2}), we have
\begin{equation}
\begin{aligned}
&|\beta_n(lt_{n_t})|^2\\
&=\frac{(n+1)g^2}{\Omega_n^2}\cos^2\left[l\pi\sqrt{\frac{\Delta^2/4+(n+1)g^2}{\Delta^2/4+(n_t+1)g^2}}\right]
+\frac{\Delta^2}{4\Omega_n^2}.
\end{aligned}
\end{equation}

In this work, we focus on the generation of a large Fock state with $n_t\gg 1$ from a coherent state $|\alpha\rangle$ with $|\alpha|^2=n_t$. Generally, the population of such a coherent state on the number state $|n\rangle$ can be negligible when $|n-n_t|\gg\sqrt{n_t}$. Consequently, in the regime of $|n-n_t|\ll\sqrt{n_t}$, we have~\cite{PRA1991Gea}
\begin{equation}\label{betan2_approx}
\begin{aligned}
&|\beta_n(lt_{n_t})|^2\\
\approx&\frac{(n+1)g^2}{\Omega_n^2}\cos^2\left[l\pi+\frac{l\pi(n-n_t)}{\Delta^2/(2g^2)+2(n_t+1)}\right]
+\frac{\Delta^2}{4\Omega_n^2}\\
=&\frac{(n+1)g^2}{\Omega_n^2}\cos^2\left[\frac{l\pi(n-n_t)}{\Delta^2/(2g^2)+2(n_t+1)}\right]+\frac{\Delta^2}{4\Omega_n^2}.
\end{aligned}
\end{equation}
By setting $l=l_k=\lfloor[\Delta^2/(4g^2)+n_t+1]/2^{k-1}\rfloor$ or $\lceil[\Delta^2/(4g^2)+n_t+1]/2^{k-1}\rceil$ for the $k$th round of evolution and measurement, GPM operator~(\ref{MP}) can then be approximately constructed.

\begin{figure}[htbp]
\centering
\includegraphics[width=0.85\linewidth]{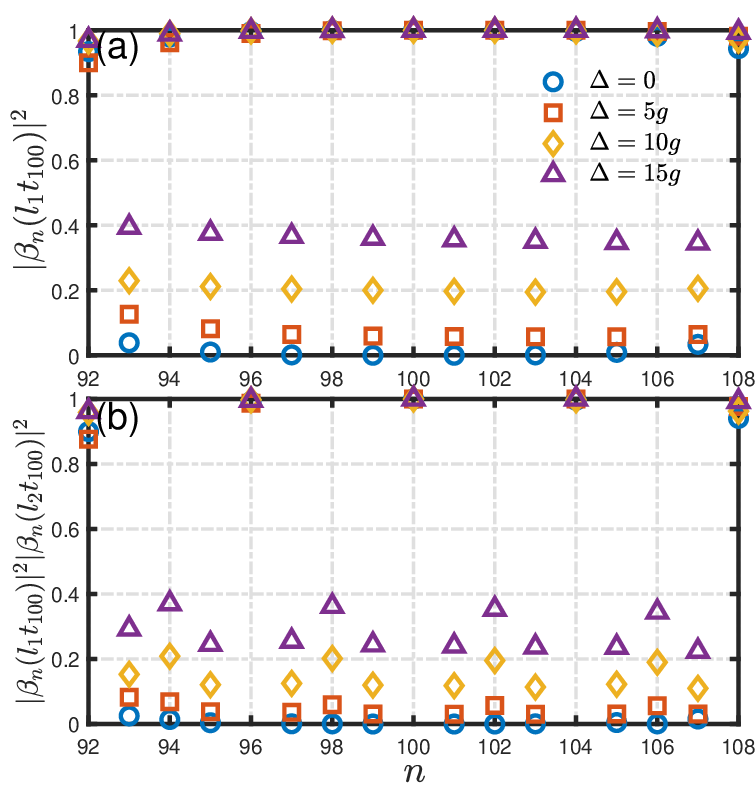}
\caption{Norm squared coefficient $\prod_{i=1}^{N}|\beta_n(l_it_{100})|^2$ as a function of the Fock index $n$ for (a) $N=1$ round and (b) $N=2$ rounds of evolution and measurement with various detunings $\Delta$. The target state is the Fock state $|n_t=100\rangle$.}\label{differ_detunings}
\end{figure}

Apparently, a nonvanishing $\Delta$ is obstructive to such a simulation of a high fidelity. Its effect can be displayed by Figs.~\ref{differ_detunings}(a) and~\ref{differ_detunings}(b), where we plot $\prod_{k=1}^{N}|\beta_n(l_kt_{n_t})|^2$ for $|n_t=100\rangle$ upon a single round and two rounds of evolution and measurement, respectively. It is shown by the blue circles in Fig.~\ref{differ_detunings}(a) that the resonant case with $\Delta=0$ gives rise to a nearly perfect GPM in the regime of $|n-n_t|\ll\sqrt{n_t}$. In particular, the target Fock state $|100\rangle$ is preserved after a single round of evolution and measurement, i.e., $|\beta_{100}(l_1t_{100})|^2=1$, and the states of odd parity are almost filtered out, e.g., $|\beta_{99}(l_1t_{100})|^2\approx|\beta_{101}(l_1t_{100})|^2\approx10^{-5}$, as expected by GPM. Note both $|\beta_{92}(l_1t_{100})|^2=0.93$ and $|\beta_{93}(l_1t_{100})|^2=0.03$ deviate slightly from the results of the standard GPM, since the approximation in Eq.~(\ref{betan2_approx}) gradually ceases to hold as $|n-n_t|$ becomes larger. In Fig.~\ref{differ_detunings}(b), it is shown that the states of parity of $2^{k=2}$ violation are almost filtered out after two rounds, e.g., $|\beta_{94}(l_2t_{100})|^2\approx0.01$, $|\beta_{98}(l_2t_{100})|^2\approx5\times10^{-5}$, $|\beta_{102}(l_2t_{100})|^2\approx6\times10^{-5}$, and $|\beta_{106}(l_2t_{100})|^2\approx4\times10^{-3}$, also as expected by GPM, while for $\Delta\neq0$ the norm squared coefficients out of the subspace of the generalized parity become nonvanishing and are verified by numerical simulation to be about $\Delta^2/(4\Omega_n^2)$. For example, in Fig.~\ref{differ_detunings}(a), $|\beta_{99}(l_1t_{100})|\approx 0.059$, $0.20$, and $0.36$ when $\Delta=5g$, $10g$, and $15g$, respectively. The residue population beyond the target state will eventually reduce the fidelity of state preparation.

Thus, $\Pi_e(\tau)$ in Eq.~(\ref{Pie}) under the resonant condition is almost the same as the nonunitary evolution operator $\mathcal{U}_g(\tau_k)$ in Eq.~(\ref{Ug}) up to irrelevant local phases:
\begin{equation}\label{PieUg}
\Pi_e(l_kt_{n_t})\approx\mathcal{U}_g(\tau_k).
\end{equation}
Thus, we set $\Delta=0$ in the following calculation unless otherwise stated. The general parity measurement $\mathcal{M}_{\rm GP}^{(N)}$ on the resonator can therefore be constructed by $N$ such rounds of evolution and measurement with stepwise halved intervals. Then $\rho_a(N)$ in Eq.~(\ref{raN}) holds except that $P(N)$ is formally replaced by
\begin{equation}\label{PeNrhoa}
 P_e(N)\approx{\rm Tr}\left[\mathcal{M}_{\rm GP}^{(N)}\rho_a(0)\mathcal{M}_{\rm GP}^{(N)}\right].
\end{equation}

\subsection{Connection between our effective GPM and cooling by measurements}\label{Connection}

Our effective GPM in Fig.~\ref{Circuit}(b) can be alternatively constructed by sequentially performing projective measurements $M_g\equiv|g\rangle\langle g|$ instead of $M_e\equiv|e\rangle\langle e|$ on the ancillary two-level system when it is initialized as $|g\rangle$. In this case, the nonunitary evolution operator for the resonator reads
\begin{equation}\label{Pig}
\Pi_g(\tau)=\langle g|U_{\rm JC}(\tau)|g\rangle=|0\rangle\langle0|
+\sum_{n=1}^{\infty}\cos(\sqrt{n}g\tau)|n\rangle\langle n|,
\end{equation}
under the resonant condition. In comparison to $\mathcal{U}_g(\tau)$ in Eq.~(\ref{Ug}), the population on the vacuum state $|0\rangle$ of the resonator is always under protection, irrespective of the joint-evolution duration $\tau$ for each round of evolution and measurement. It will generally decrease the final fidelity of the target state.

Based on this special property of $\Pi_g(\tau)$, many protocols have been proposed for cooling the resonators initially in a finite-temperature state~\cite{PRB2011Li,PRA2021Yan} and generating the Fock state superposition $c_0|0\rangle+c_n|n\rangle$ from a coherent state~\cite{PRA2024Zhang} by selective measurements. Here for generating a large Fock state $|n_t\gg 1\rangle$ from the coherent state $|\alpha=\sqrt{n_t}\rangle$, the initial population on the ground state $|0\rangle$ is negligible and thus does not significantly impact the final fidelity of the target Fock state. Then the duration of the joint evolution can be optimized as
\begin{equation}\label{taug}
    \tau=\frac{l\pi}{\sqrt{n_t}g}
\end{equation}
to have $|\langle n_t|\Pi_g(\tau)|n_t\rangle|=1$, where $l\in \mathbb{N}$. By substituting Eq.~(\ref{taug}) for Eq.~(\ref{Pig}), the nonunitary operator in the regime of $n_t\gg 1$ and $|n-n_t|\ll n_t$ can be approximated in the same way as Eq.~(\ref{betan2_approx}):
\begin{equation}
\begin{aligned}
 \Pi_g\left(\frac{l\pi}{\sqrt{n_t}g}\right)=&\sum_n\cos\left(l\pi\sqrt{\frac{n}{n_t}}\right)|n\rangle\langle n|\\
 \approx&\sum_n\cos\left(l\pi+l\pi\frac{n-n_t}{2n_t}\right)|n\rangle\langle n|\\
 =&\sum_n(-1)^l\cos\left(l\pi\frac{n-n_t}{2n_t}\right)|n\rangle\langle n|.
\end{aligned}
\end{equation}
GPM can therefore be approximately constructed by setting $l=l_k=\lfloor n_t/2^{k-1}\rfloor$ or $\lceil n_t/2^{k-1}\rceil$ for the $k$th round of evolution and measurement. Equation~(\ref{PieUg}) is then modified as $\Pi_g(l_kt_{n_t-1})\approx\mathcal{U}_g(\tau_k)$.

In a short summary, our GPM protocol for generating large Fock states based on $\Pi_g(\tau)$ carries a similar idea as the measurement-induced cooling protocols, except for the setting of the joint-evolution period in each round.

\section{Large Fock state generation}\label{LargeFock}

By leveraging the approximated generalized parity measurement constructed by the circuit in Fig.~\ref{Circuit}(b), we can efficiently generate a large number state $|n_t\rangle$ of the resonator using merely projective measurements on the ancillary TLS, with no additional gate operations on either resonator or ancillary qubit.

\begin{figure}[htbp]
\centering
\includegraphics[width=0.85\linewidth]{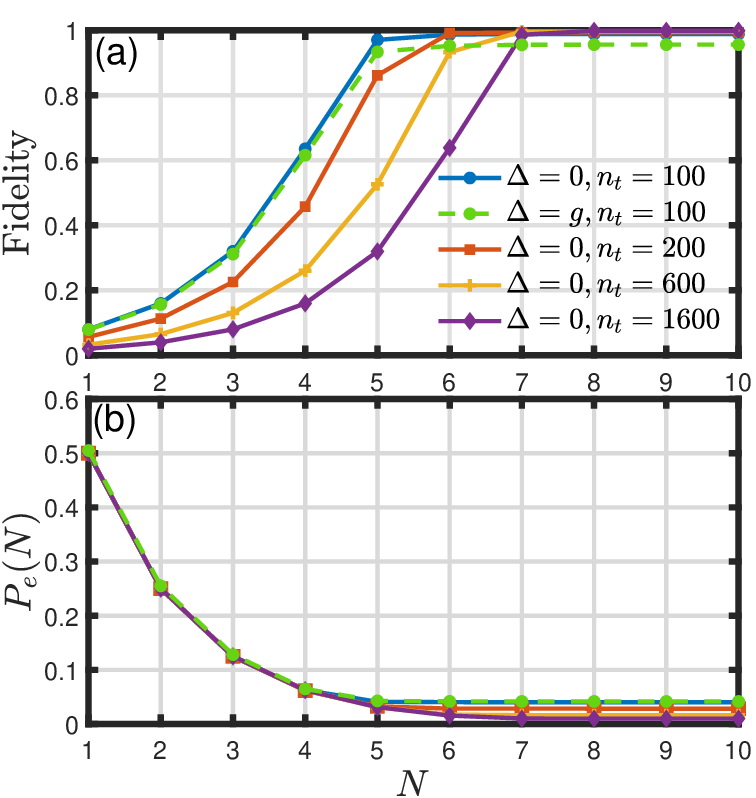}
\caption{(a) Fidelity and (b) success probability as functions of the number $N$ of rounds of free evolution and measurement for various target Fock states $|n_t\rangle$.}\label{Fock_gene}
\end{figure}

In Fig.~\ref{Fock_gene}(a), we present the fidelity $F=\langle n_t|\rho_a(N)|n_t\rangle$ of various target Fock states $|n_t\rangle$ as a function of the measurement number $N$. The results indicate that the fidelities rapidly converge to near unit in less than ten rounds of measurements across a wide range of $n_t$. In particular, $N=6$ and $8$ measurements are sufficient to generate $|n_t=100\rangle$ and $|1600\rangle$ with $F\geq0.99$, respectively. Figure~\ref{Fock_gene}(b) demonstrates the relevant success probability $P_e(N)$ of finding the qubit at its excited state as a function of $N$. The steady-state success probabilities for the generated Fock states with $n_t=100$, $200$, $600$, and $1600$ are $P_e\approx0.04$, $0.028$, $0.016$, and $0.009$, respectively. In the absence of decoherence, they are determined by the overlap between the initial coherent state and the target Fock state $P_e(N)\rightarrow|\langle n_t|\alpha\rangle|^2=e^{-n_t}n_t^{n_t}/n_t!$. Any deviation of $|\alpha|^2$ from $n_t$ therefore causes a certain reduction of the steady-state success probability. Using the Stirling formula $n_t!\approx e^{-n_t}n_t^{n_t}\sqrt{2\pi n_t}$, we have $P_e(\infty)\approx 1/\sqrt{2\pi n_t}$ that varies with a small rate for $n_t\gg 1$. Note the variance of the population distribution of the coherent state scales as $\langle\Delta n\rangle\sim\sqrt{n_t}$. It implies that our protocol is insensitive to the small deviation between the initial average occupation and the target photon number $||\alpha|^2-n_t|<\sqrt{n_t}$ when $n_t$ is sufficiently large. Moreover, during the first four rounds, the success probability descends roughly as $2^{-N}$, irrespective of $n_t$. It can be understood by the fact that for a coherent state with $|\alpha|^2=n_t\gg1$, the population distribution is an approximately Gaussian function centered around $n_t$~\cite{JMO1989Barnett}, which is given by
\begin{equation}
|\langle n|\alpha\rangle|^2\approx\sqrt{\frac{1}{2\pi n_t}}\exp\left[-\frac{(n-n_t)^2}{2n_t}\right].
\end{equation}
After $N$ rounds of measurements, the success probability in Eq.~(\ref{PeNrhoa}) turns out to be
\begin{equation}
\begin{aligned}
P_e(N)\approx&{\rm Tr}\left[\mathcal{M}_{\rm GP}^{(N)}|\alpha\rangle\langle\alpha|\mathcal{M}_{\rm GP}^{(N)}\right]\\
=&\sqrt{\frac{1}{2\pi n_t}}\sum_{j\geq\lfloor -n_t/2^N\rfloor }\exp\left(-\frac{j^22^{2N}}{2n_t}\right),\label{PeN}
\end{aligned}
\end{equation}
due to Eq.~(\ref{MP}). For $n_t2^{1-2N}\gg1$, the summation over $j$ can be approximated by the Gaussian integral
\begin{equation}
\begin{aligned}
&\sum_{j\geq\lfloor -n_t/2^N\rfloor }\exp\left(-\frac{j^2}{n_t2^{1-2N}}\right)\\
&\approx\int_{-\infty}^{+\infty}\exp\left(-\frac{x^2}{n_t2^{1-2N}}\right)dx=\sqrt{\pi n_t2^{1-2N}}.\label{Gauss_int}
\end{aligned}
\end{equation}
By inserting Eq.~(\ref{Gauss_int}) into Eq.~(\ref{PeN}), it is found that the success probability $P_e(N)\approx2^{-N}$ in the first several rounds, which is independent of the target state.

The performance of our protocol in the presence of a nonvanishing detuning is shown by the green dashed line marked with circles in Fig.~\ref{Fock_gene}(a). The output fidelity of the target Fock state is about $95\%$ after ten rounds of evolution and measurement under a moderate detuning $\Delta=g$, which is the accumulated effect indicated by Fig.~\ref{differ_detunings}. It is consistent with our analysis that $\Delta\neq0$ yields residue populations outside the subspace of general parity, which cannot be eliminated by our approximated GPM. In Fig.~\ref{Fock_gene}(b), the behavior of the success probability for $\Delta\neq0$ is almost identical to the resonant case.

\begin{figure}[htbp]
\centering
\includegraphics[width=0.85\linewidth]{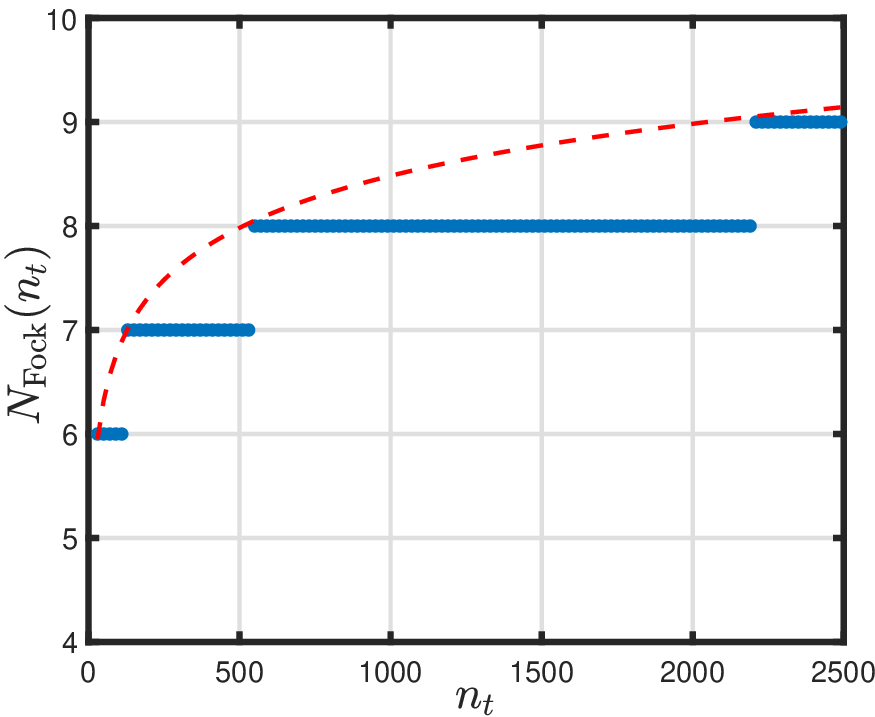}
\caption{Number of evolution-measurement rounds for generating a Fock state $|n_t\rangle$ with a fidelity exceeding $98\%$ as a function of the target excitation number $n_t$. The red dashed line is the fitting curve $N_{\rm  Fock}(n_t)\approx\log_2(\sqrt{n_t})+3.5$. }\label{HO_steps}
\end{figure}

Figure~\ref{HO_steps} plots the number of evolution-measurement rounds $N_{\rm Fock}(n_t)$ required to prepare a Fock state $|n_t\rangle$ with a fidelity exceeding $98\%$. The blue lines are obtained by the numerical results. The red dashed line indicates the fitting curve $N_{\rm Fock}(n_t)\approx\log_2(\sqrt{n_t})+3.5$, showing that $N_{\rm Fock}(n_t)$ roughly follows a logarithmical function of the square root of the target excitation number $n_t$. It confirms that our protocol is almost exponential in efficiency.

In comparison to the existing GPM protocols in Fig.~\ref{Circuit}(a) using the dispersive coupling between an ancillary TLS and the target resonator, our protocol in Fig.~\ref{Circuit}(b) offers at least three advantages: (1) The dispersive-coupling approach employs two Hadamard gates per round of evolution and measurement, that is more complex in performance and induces extra gate infidelity. (2) The total running time of our protocol is $T_{\rm JC}\propto\sqrt{n_t}/g$, whereas the time consumption of the dispersive-coupling protocol is $T_{\rm dis}\propto1/\chi$. It is straightforward to find that the former is faster than the latter as long as $n_t<g^2/\chi^2$. In the current circuit QED platforms with a typical dispersive-coupling strength $\chi\approx 1$ MHz between the superconducting qubit and the microwave cavity mode and the exchange coupling strength $g\approx 100$ MHz~\cite{Nature2008SHo,PRL2017Wang,PRApplied2017Brecht}, our protocol holds a low cost of running time for generating Fock states within $n_t<10\ 000$. (3) In both protocols, the period of each round is set as one-half of the last round. Yet one has to modify the detunings $\Delta$ for different target states $|n_t\rangle$ in the dispersive-coupling protocol.

In a practical experiment, decoherence and dephasing caused by the surrounding environment would degrade the attainable fidelity of any target state. Under a zero-temperature environment, the time evolution of the composite system can be described by the Lindblad master equation
\begin{equation}
\begin{aligned}
\dot{\rho}(t)=&-i[H_{\rm JC},\rho(t)]+\kappa\mathcal{L}[a]\rho(t)\\
&+\gamma\mathcal{L}[|g\rangle\langle e|]\rho(t)+\gamma_{\phi}\mathcal{L}[|e\rangle\langle e|]\rho(t),
\end{aligned}
\end{equation}
where $\gamma$ and $\kappa$ are the decay rates of the ancillary qubit and target resonator, respectively, $\gamma_{\phi}$ is the dephasing rate of the qubit, and $\mathcal{L}[o]$ is the Lindblad superoperator defined as $\mathcal{L}[o]\rho\equiv o\rho o^{\dagger}-1/2(o^{\dagger}o\rho+\rho o^{\dagger}o)$. In a 3D circuit QED platform, the decay rate of the microwave cavity $\kappa\approx 1$--$10$ kHz. The qubit relaxation time and dephasing time are about $T_1\approx T_{\phi}\approx 10$--$100$ $\mu$s. Then $\gamma\approx\gamma_{\phi}\approx 10$--$100$ kHz for a transmon qubit~\cite{Nature2008SHo,PRL2017Wang,PRApplied2017Brecht}. In the numerical simulation, we assume $\gamma=\gamma_{\phi}$ and set the resonant and dispersive coupling strengths as $g=100$ MHz and $\chi=2$ MHz, respectively.

\begin{figure}[htbp]
\centering
\includegraphics[width=0.85\linewidth]{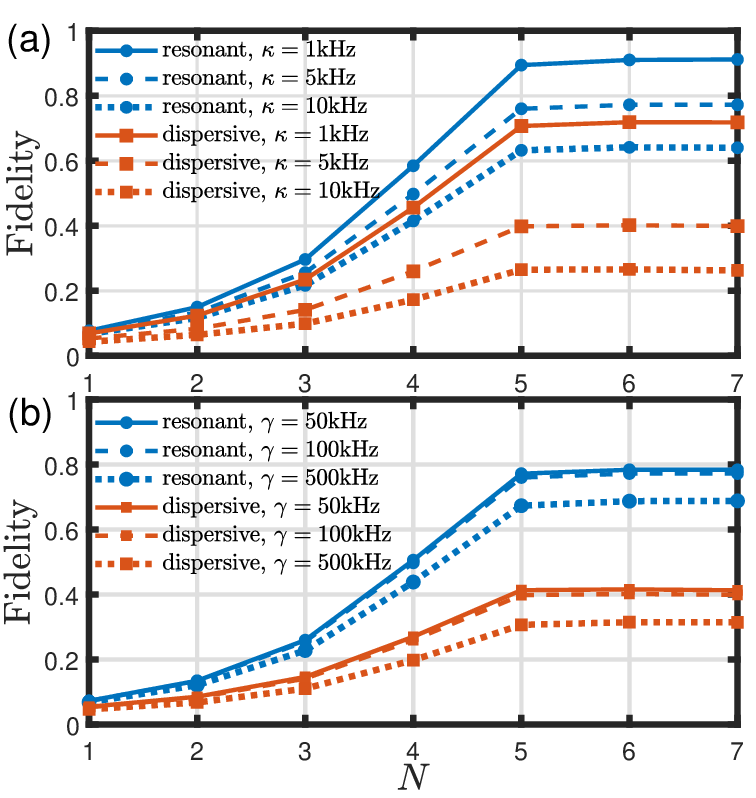}
\caption{Comparison of the resonant exchange coupling protocol [Fig.~\ref{Circuit}(b)] and the dispersive-coupling protocol [Fig.~\ref{Circuit}(a)] for GPM in generating Fock state $|n_t=100\rangle$ (a) under various cavity decay rates with $\gamma=\gamma_{\phi}=100$ kHz and (b) under various qubit decoherence rates ($\gamma=\gamma_\phi$) with $\kappa=5$ kHz. Here $g=100$ MHz and $\chi=2$ MHz.}\label{HO_differ_kappa}
\end{figure}

In Fig.~\ref{HO_differ_kappa}(a), we compare the performance of our protocol for the approximated GPM in which the target resonator is exchangeably coupled to an ancillary qubit with that of the existing dispersive-coupling protocol under various cavity decay rates. It is found that our resonant-coupling protocol outperforms the dispersive one and exhibits a significantly higher fidelity under the same condition. After $N=7$ measurements, the fidelities of the former reach $91$, $77$, and $64\%$ while the fidelities of the latter are about $71$, $40$, and $26\%$ for $\kappa=1$, $5$, and $10$ kHz, respectively. Similar advantage of our resonant-coupling protocol can be confirmed under various qubit decoherence rates as shown in Fig.~\ref{HO_differ_kappa}(b), where the resonator's decay rate is fixed as $\kappa=5$ kHz. It is found that the dispersive-coupling protocol becomes inapplicable since all the fidelities are under $50\%$, while the fidelity of our protocol is still above $68\%$ even using a low-quality ancillary qubit with $\gamma=\gamma_\phi=500$ kHz. The outstanding performance of our protocol mainly stems from the reduced time consumption of the generation procedure. The output fidelity nearly saturates in both protocols with $N=5$ rounds of measurements for generating the Fock state $|n_t=100\rangle$. Nevertheless, it takes $T_{\rm JC}\approx\sum_{k=1}^5\sqrt{n_t+1}\pi/(2^{k-1}g)\approx 612$ ns in our protocol, while $T_{\rm dis}=\sum_{k=1}^5\pi/(2^{k-1}\chi)\approx 3043$ ns in the dispersive-coupling protocol.

\section{Preparing the Dicke state via GPM}\label{preparingDicke}

Another advantage of our circuit model in Fig.~\ref{Circuit}(b) is its versatility in the state generation. Under certain conditions, an effective generalized parity measurement can be constructed to prepare the Dicke state of a large excitation number, which is one of the resource states for quantum metrology~\cite{PRA2012Toth,NJP2014Zhang,NJP2015Apellaniz,RMP2018Pezz,PRA2020Hakoshima}. Numerous protocols have been proposed for generating the Dicke states in the systems of several qubits~\cite{PRL2003Duan,PRA2008Ionicioiu,PRALamata2013,PRA2017Wu,PRA2023Stojanovi}. However, extending these methods to a large scale remains a significant challenge due to the requirement of the individual addressing the probe qubits or the inner coupling among them, which is costly in a large spin-ensemble, e.g., the diamond nitrogen-vacancy centers~\cite{NC2014Dolde}. In sharp contrast, our protocol needs merely to locally address and readout the ancillary system, e.g., the central spin in the spin-star model.

\subsection{Effective GPM in the spin-star model}\label{GPMDicke}

Consider a spin-star model consisting of a central spin-$1/2$ coupled to a spin ensemble containing $M$ identical spin-$1/2$'s via a Heisenberg $XY$ interaction of identical coupling strength~\cite{PRB2007Jing,PRA2004Hutton,PRB2004Breuer,PRA2019Radhakrishnan}. The full Hamiltonian reads
\begin{equation}
H=\frac{\omega_c}{2}\sigma_c^z+\frac{\omega_b}{2}\sum_{j=1}^M\sigma^z_j
+g\sum_{j=1}^M(\sigma_c^x\sigma_j^x+\sigma_c^y\sigma_j^y),
\end{equation}
where $\sigma^k$'s, $k=x,y,z$, are the Pauli operators along various directions and $g$ is the spin-spin coupling strength. $\omega_c$ and $\omega_b$ are the energy splitting of the center spin and the surrounding spins, respectively. By rewriting the interaction Hamiltonian in terms of the transition operators $\sigma^+=|e\rangle\langle g|$ and $\sigma^-=|g\rangle\langle e|$ and transforming the full Hamiltonian to the rotating frame with respect to $H_0=\omega_b/2(\sigma^z_c+\sum_{j=i}^M\sigma^z_j)$, we have
\begin{equation}
H'=\frac{\Delta}{2}\sigma^z_c+2g\left(\sigma_c^+J_-+\sigma_c^-J_+\right),
\end{equation}
where $\Delta=\omega_c-\omega_b$ is the detuning between the center spin and the spins in the spin ensemble. $J_\pm\equiv\sum_{j=1}^M\sigma^\pm_j$ are the collective angular momentum operators. We assume that the state of the spin ensemble is restricted to the permutation symmetric subspace, i.e., the subspace with the total spin $J=M/2$. Any state $|J,m_t\rangle$ with $-M/2<m_t<M/2$ in that subspace could be regarded as a general Dicke state. Such a model was used for generating the Dicke state via repeated exchange of excitation between the central spin and the spin ensemble, and each round consists of a flipping on the center spin and a half Rabi oscillation of the composite system~\cite{PRA2020Hakoshima}. Staring from a polarized state $|J,\pm J\rangle$, it is extremely hard to generate the ``center'' Dicke state of a large spin ensemble $|J=M/2\gg 1,0\rangle$.

The collective angular momentum operator $J_\pm$ in this subspace can be rewritten as
\begin{equation}
 J_\pm=\sum_{m=-J}^J\sqrt{J(J+1)-m(m\pm 1)}|J,m\pm 1\rangle\langle J,m|.
\end{equation}
The center spin is supposed to be initially prepared as one of its energy eigenstates $|i\rangle$, $i=e,g$, and the initial state of the spin ensemble is $\rho_b(0)$. Then the separable initial state of the composite system is
\begin{equation}
 \rho(0)=|i\rangle\langle i|\otimes\rho_b(0).
\end{equation}
Similar to the generation of large Fock states, after a period of free evolution $U(\tau)=\exp(-iH'\tau)$, we perform a projective measurement $M_i=|i\rangle\langle i|$ on the center spin. The state of the spin ensemble becomes
\begin{equation}
\rho_b(1)=\frac{V_i(\tau)\rho_b(0)V_i^{\dagger}(\tau)}{P_i(1)},
\end{equation}
where $P_i(1)={\rm Tr}[V_i(\tau)\rho_b(0)V_i^{\dagger}(\tau)]$ is the success probability of finding the center spin in state $|i\rangle$, and $\rho_b(k)$ indicates the density matrix of the spin ensemble after $k$ rounds of measurements. The effective nonunitary evolution operator $V_i(\tau)=\langle i|U(\tau)|i\rangle$ reads
\begin{equation}
V_i(\tau)=\sum_{m=-J}^J\lambda_m^{(i)}(\tau)|J,m\rangle\langle J,m|,
\end{equation}
where the coefficients $\lambda_m^{(i)}(\tau)$'s are given by
\begin{equation}
\lambda_m^{(i)}(\tau)=e^{-i\Delta \tau/2}\left[\cos\left(\Lambda^{(i)}_m\tau\right)+i\frac{\Delta}{2\Lambda^{(i)}_m}\sin\left(\Lambda_m^{(i)}\tau\right)\right],
\end{equation}
with
\begin{equation}
\Lambda_m^{(i)}=\sqrt{\Delta^2/4+4g^2[J(J+1)-m(m\pm1)]}.
\end{equation}
Here and in the following $i=e$ and $g$ correspond to the upper and lower signs, respectively. To hold the population on the target Dicke state $|J,m_t\rangle$, the relevant coefficients have to satisfy $|\lambda_{m_t}^{(i)}(\tau)|=1$, which renders the optimized interval of each round
\begin{equation}
\tau=\xi t^{(i)}_{m_t}, \quad t^{(i)}_{m_t}\equiv\frac{\pi}{\Lambda_{m_t}^{(i)}}
\end{equation}
with $\xi\in\mathbb{N}$. Under the resonant condition $\Delta=0$, the nonunitary evolution operator for each round of evolution and measurement turns out to be
\begin{align}
&V_i\left(\xi t_{m_t}^{(i)}\right)=\nonumber\\ \label{Vemt}
&\sum_{m=-J}^J\cos\left[\xi\pi\sqrt{\frac{J(J+1)-m(m\pm1)}{J(J+1)-m_t(m_t\pm1)}}
\right]|J,m\rangle\langle J,m|.
\end{align}

Suppose the spin ensemble is initially prepared in the tensor product of spin coherent states, i.e., $\rho_b(0)=|\psi_b(0)\rangle\langle\psi_b(0)|$, where
\begin{align}
|\psi_b(0)\rangle&=\left[\cos\left(\frac{\phi}{2}\right)|g\rangle+\sin\left(\frac{\phi}{2}\right)|e\rangle\right]^{\otimes M}\nonumber\\ &=\sum_{m=-J}^Jp_M(m)|J,m\rangle,\\
p_M(m)&=\sqrt{\binom{M}{m+M/2}}\cos^{\frac{M}{2}-m}\left(\frac{\phi}{2}\right)\sin^{\frac{M}{2}+m}\left(\frac{\phi}{2}\right).
\end{align}
The overlap between the initial state and target Dicke state is maximized when $\phi=\arccos(-2m_t/M)$~\cite{PRL2025Nagib}. In this case, the populations on the deviated components $|J,m\rangle$ with $|m-m_t|\gg\sqrt{M}/2$ are negligible, since the standard deviation of $|p_M(m)|^2$ is $\sqrt{M}/2$~\cite{PRA2021Wang}. Thus, we can only consider the Dicke state components near the target Dicke state $|J,m_t\rangle$.

For simplicity, we now focus on the generation of the center Dicke state $|J,0\rangle$. The other Dicke state with $m_t/M\ll1$ can be treated in a similar way. Thus, the nonunitary evolution operators in Eq.~(\ref{Vemt}) can be approximated as
\begin{equation}\label{Veg}
\begin{aligned}
&V_i\left(\xi t_{0}^{(i)}\right)=\sum_{m=-J}^J\cos\left[\xi\pi\sqrt{1-\frac{m(m\pm 1)}{J(J+1)}}\right]|J,m\rangle\langle J,m|\\
&\approx \sum_{m=-J}^J\cos\left[\xi\pi-\xi\pi\frac{m(m\pm 1)}{2J(J+1)}\right]|J,m\rangle\langle J,m|\\
&=\sum_{m=-J}^J(-1)^\xi\cos\left[\xi\pi\frac{m(m\pm 1)}{2J(J+1)}\right]|J,m\rangle\langle J,m|.
\end{aligned}
\end{equation}
By setting $\xi=\xi_k=\lfloor J(J+1)/2^k\rfloor$ or $\lceil J(J+1)/2^k\rceil$ for the $k$th round of evolution and measurement, the spin ensemble is projected onto the subspace spanned by the set of Dicke states $|J,m\rangle$ with $|m(m\pm 1)/2^{N+1}|\in \mathbb{N}$ after $N$ rounds, which is equivalent to running an effective generalized parity measurement on the spin ensemble
\begin{equation}\label{rhobN}
\rho_b(N)=\frac{\mathcal{P}_i^{(N)}\rho_b(0)\mathcal{P}_i^{(N)}}{P(N)}.
\end{equation}
The effective GPM operator reads
\begin{equation}
\begin{aligned}
\mathcal{P}_i^{(N)}\equiv&\sum_j\Big(|J,2^{N+1}j\rangle\langle J,2^{N+1}j|\\
&+|J,2^{N+1}j\mp 1\rangle\langle J,2^{N+1}j\mp 1|\Big). \label{PiN}
\end{aligned}
\end{equation}
Similar to the Fock state generation, the number of measurement rounds for generating a Dicke state near the center Dicke state $|J,0\rangle$ roughly scales as $N_{\rm Dicke}(M)\sim \log_2(\sqrt{M})$, since the standard deviation of $|p_M(m)|^2$ is $\sqrt{M}/2$.

\subsection{Dicke state generation}\label{Dickeresult}

\begin{figure}[htbp]
\centering
\includegraphics[width=0.85\linewidth]{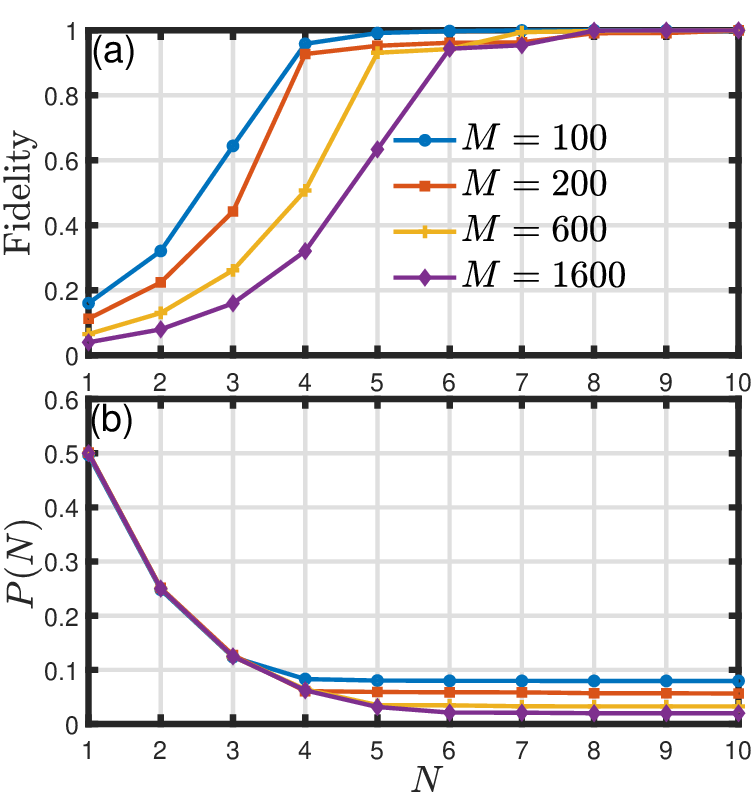}
\caption{Dependence of (a) fidelity of the center Dicke state $|J,0\rangle$ and (b) success probability on the number of rounds $N$ for various spin numbers $M$ in the spin ensemble.}\label{Dicke_differ_M}
\end{figure}

It is straightforward to find from Eq.~(\ref{PiN}) that the target Dicke state $|J,0\rangle$ cannot be distinguished from  $|J,1\rangle$ ($|J,-1\rangle$) by only using the nonunitary evolution operator $V_g(\tau)$ [$V_e(\tau)$] in Eq.~(\ref{Veg}). Thus we can take a hybrid strategy. In the first round, we prepare the center spin in the excited state $|e\rangle$ and perform the projective measurement $M_e=|e\rangle\langle e|$ on the center spin after a period of $\tau=\xi_1t^{(e)}_0$ to filter out the population on $|J,1\rangle$. And then we use a pulse to flip the center spin from $|e\rangle$ to $|g\rangle$ and begin to perform rounds of the evolution and measurement with the duration $\tau=\xi_{k-1}t^{(g)}_0$ and the projective measurement operator $M_g=|g\rangle\langle g|$ in the subsequent $k$th rounds with $k\geq2$. Therefore, only an extra single gate is required in our protocol, which is substantially simpler than the standard QPE algorithm~\cite{PRA2021Wang,PRL2024Piroli}.

In Figs.~\ref{Dicke_differ_M}(a) and \ref{Dicke_differ_M}(b), we present the fidelity $F=\langle J,0|\rho_b(N)|J,0\rangle$ and the success probability $P(N)$ for generating the center Dicke state $|J=M/2,0\rangle$ with various spin numbers $M$, respectively. For a wide range from $M=100$ to $1600$, we have $F>0.99$ by eight rounds of evolution and measurement. Similar to Fock state generation, the asymptotic value of success probabilities is determined by the overlap between the target state and the initial product state. Using the Stirling formula, it is found to be
\begin{equation}
P(N)\rightarrow|p_M(0)|^2=\binom{M}{M/2}2^{-M}\approx \sqrt{\frac{2}{\pi M}}.
\end{equation}

\begin{figure}[htbp]
\centering
\includegraphics[width=0.85\linewidth]{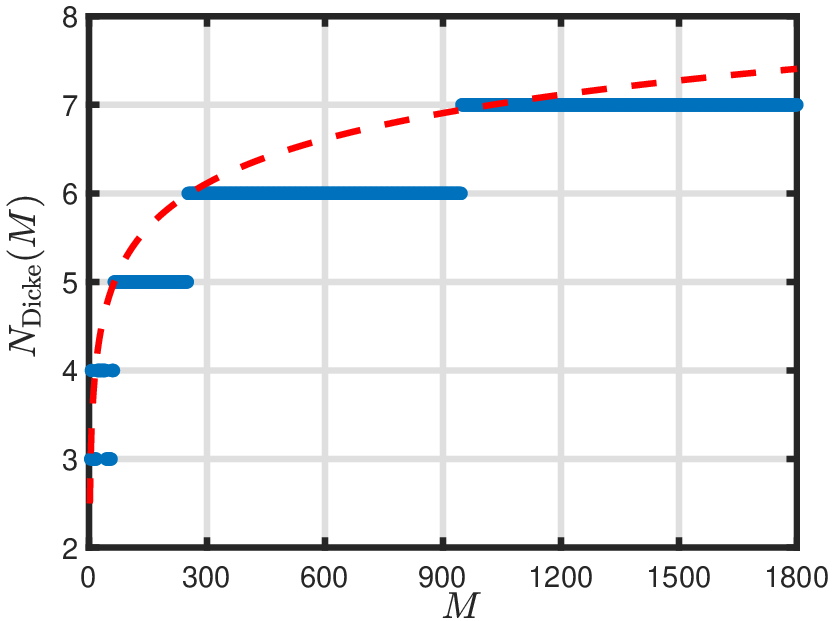}
\caption{Number of evolution-measurement rounds for generating the center Dicke state with a fidelity exceeding $90\%$ as a function of the size of spin ensemble $M$. The red dashed line represents the fitting curve $N_{\rm  Dicke}(M)=\log_2(\sqrt{M})+2$.}\label{Dicke_steps}
\end{figure}

Again, due to a similar reason as in the Fock state generation, the success probability also exhibits an $M$-independent behavior in the first few rounds. For $M\gg 1$, the population distribution in the subspace $|p_M(m)|^2$ is found to be approximately a Gaussian function: $|p_M(m)|^2\approx \sqrt{2/(\pi M)}\exp(-2m^2/M)$. Owing to the fact that
\begin{equation}
\begin{aligned}
&{\rm Tr}\left[\mathcal{P}_i^{(N)}|\psi_b(0)\rangle\langle\psi_b(0)|\mathcal{P}_i^{(N)}\right]\\
=&\sqrt{\frac{2}{\pi M}}\sum_j\left[\exp\left(-\frac{j^2}{M2^{-3-2N}}\right)+\exp\left(-\frac{(j\pm 1)^2}{M2^{-3-2N}}\right)\right]\\
\approx&2\sqrt{\frac{2}{\pi M}}\int_{-\infty}^{+\infty}dx\exp\left(-\frac{x^2}{M2^{-3-2N}}\right)=2^{-N}
\end{aligned}
\end{equation}
under the condition $M\gg2^{3+2N}$, the success probability is about $2^{-N}$. Figure~\ref{Dicke_steps}  presents the number $N_{\rm Dicke}(M)$ of rounds required by our protocol for generating Dicke state $|J=M/2,0\rangle$ as a function of spin number $M$. The red-dashed line is a fitting curve with $N_{\rm Dicke}=\log_2(\sqrt{M})+2$, which confirms an approximate logarithmic scaling supported by GPM or QPE.

\begin{figure}[htbp]
\centering
\includegraphics[width=0.85\linewidth]{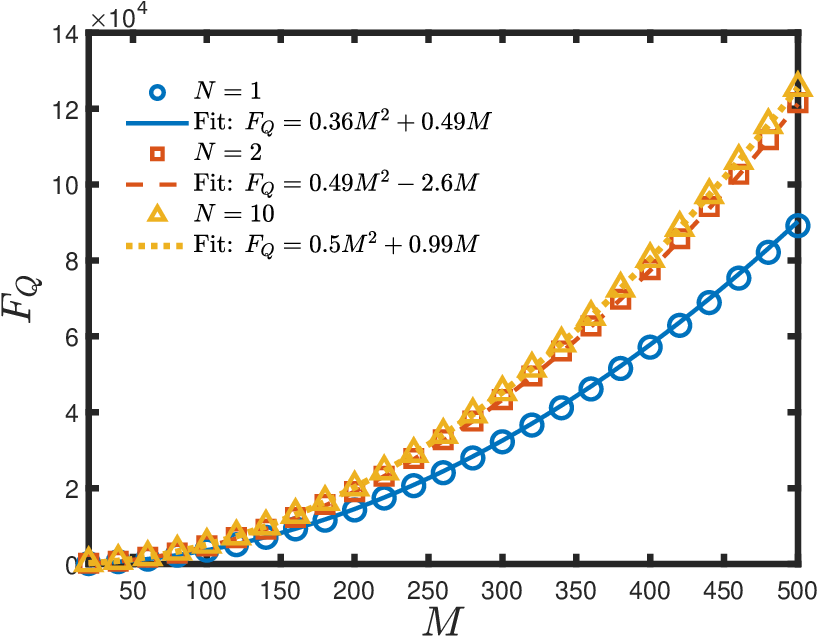}
\caption{Quantum Fisher information about estimating the phase $\theta$ along the $x$ axis as a function of the spin-probe size $M$ after $N$ rounds of evolution and measurement.}\label{FQ_differ_N}
\end{figure}

To use the Dicke state generated by our protocol in quantum metrology, it is not necessary to increase the fidelity to an extremely high level through a lot of repeated measurements. Here we calculate the quantum Fisher information (QFI)~\cite{PRL1994Braunstein} about the phase estimation along the $x$ axis encoded by $R_x(\theta)=\exp(-i\theta J_x)$ to the probe the spin ensemble. In our protocol, the spin ensemble remains in a pure state. Consequently, the output QFI for estimating $\theta$ is given by $F_Q=4\langle \psi_b(N)|J_x^2|\psi_b(N)\rangle-4|\langle \psi_b(N)|J_x|\psi_b(N)\rangle|^2$. Here $|\psi_b(N)\rangle$ is the output state of the spin ensemble after $N$ rounds of evolution and measurement, which is obtained by Eq.~(\ref{rhobN}). For an ideal center Dicke state $|J=M/2,0\rangle$ of a spin ensemble containing $M$ spins, QFI is found to be $F_Q(M)=M^2/2+M$~\cite{PRA2012Toth}.

In Fig.~\ref{FQ_differ_N}, we plot QFI as a function of the ensemble size $M$ after $N=1$ (blue circles), $N=2$ (red squares), and $N=10$ (yellow triangles) rounds of evolution and measurement. In each case, the data are well described by a quadratic behavior $F_Q(M)\approx aM^2+bM$. It is interesting to find that QFI exhibits a significantly faster convergence than the fidelity in Fig.~\ref{Dicke_differ_M}. For $N=2$, the factor $a$ for the square term has already approached the ideal value $1/2$ for the Heisenberg limit in metrology precision.

\section{Conclusion}\label{conclusion}

In summary, we presented a measurement-based protocol for the efficient preparation of nonclassical states with a large number of excitations in bosonic and spin-ensemble systems via the resonant exchange interaction between the ancillary qubit and the target system and the projective measurement on the qubit. By repeatedly performing the evolution-measurement rounds with carefully tailored duration of the joint evolution in each round, one can construct an effective generalized-parity measurement operator that projects the target system to large Fock states or large Dicke states. Our analysis proves that the required number of rounds grows logarithmically with the square root of the target excitation number, which is as efficient as those protocols supported by the quantum phase estimation algorithm or Ramsey circuit. In contrast to the protocols based on the dispersive interaction Hamiltonian, our protocol is featured with a much smaller number of operations. Numerical simulation adopting realistic cavity and qubit decoherence rates confirms that large Fock states can be generated with a high fidelity in six measurement cycles with the excitation number over $100$. Moreover, our protocol provides a simple method to generate an approximate Dicke state, which can achieve a Heisenberg scaling behavior in quantum Fisher information for phase estimation with only two rounds of measurement. This work demonstrates that the approximated generalized parity measurement could become a powerful and practical tool for quantum state engineering.

\section*{Acknowledgment}

We acknowledge grant support from the National Natural Science Foundation of China (Grant No. U25A20199) and the ``Pioneer'' and ``Leading Goose'' Research and Development programs of Zhejiang Province (Grant No. 2025C01028).

\bibliographystyle{apsrevlong}
\bibliography{ref}

\end{document}